%% file: bare_conf_NDSS2024.tex
\documentclass[conference]{IEEEtran}
\usepackage{tabularx}
\usepackage{multirow}
\usepackage{multicol}
\usepackage{booktabs}
\usepackage{array}
\usepackage{marvosym}
\usepackage{bbding}
\usepackage{amssymb}
\usepackage{mathtools}
\usepackage{cuted}
\usepackage[ruled,linesnumbered]{algorithm2e}

\usepackage{cite}

\ifCLASSINFOpdf
  \usepackage[pdftex]{graphicx}
  \graphicspath{{../pdf/}{../jpeg/}}
  \DeclareGraphicsExtensions{.pdf,.jpeg,.png}
\else
  \usepackage[dvips]{graphicx}
  \graphicspath{{../eps/}}
  \DeclareGraphicsExtensions{.eps}
\fi
\usepackage{algorithmic}

\usepackage{mdwmath}
\usepackage{mdwtab}

\usepackage{eqparbox}

\usepackage{caption}
\usepackage[font=footnotesize]{subfig}
\usepackage{fixltx2e}

\usepackage{stfloats}

\usepackage{url}

\usepackage{float}

\usepackage{svg}

\begin{document}
%
\title{BioDeepHash: Mapping Biometrics into a Stable Code}



%
\IEEEoverridecommandlockouts
\author{\IEEEauthorblockN{Baogang Song\IEEEauthorrefmark{1},
Dongdong Zhao\thanks{\Letter $\enspace$ Corresponding author: zdd@whut.edu.cn.}\IEEEauthorrefmark{1},
Jiang Yan\IEEEauthorrefmark{1}, 
Huanhuan Li\IEEEauthorrefmark{2} and
Hao Jiang\IEEEauthorrefmark{3}}
\IEEEauthorblockA{\IEEEauthorrefmark{1}School of Computer Science and Artificial Intelligence, Wuhan University of Technology}
\IEEEauthorblockA{\IEEEauthorrefmark{2}School of Computer Science, China University of Geosciences}
\IEEEauthorblockA{\IEEEauthorrefmark{3}School of Computer Science and
Technology, Anhui University}}



\maketitle

\begin{abstract}
    With the wide application of biometrics, more and more attention has been paid to the security of biometric templates. However most of existing biometric template protection (BTP) methods have some security problems, e.g. the problem that protected templates leak part of the original biometric data (exists in Cancelable Biometrics (CB)), the use of error-correcting codes (ECC) leads to decodable attack, statistical attack (exists in Biometric Cryptosystems (BCS)), the inability to achieve revocability (exists in methods using Neural Network (NN) to learn pre-defined templates), the inability to use cryptographic hash to guarantee strong security (exists in CB and methods using NN to learn latent templates). In this paper, we propose a framework called BioDeepHash based on deep hashing and cryptographic hashing to address the above four problems, where different biometric data of the same user are mapped to a stable code using deep hashing instead of predefined binary codes thus avoiding the use of ECC. An application-specific binary string is employed to achieve revocability.  Then cryptographic hashing is used to get the final protected template to ensure strong security. Ultimately our framework achieves not storing any data that would leak part of the original biometric data. We also conduct extensive experiments on facial and iris datasets. Our method achieves an improvement of 10.12$\%$ on the average Genuine Acceptance Rate (GAR) for iris data and 3.12$\%$ for facial data compared to existing methods. In addition, BioDeepHash achieves extremely low False Acceptance Rate (FAR), i.e. 0$\%$ FAR on the iris dataset and the highest FAR on the facial dataset is only 0.0002$\%$.
\end{abstract}



%

\input{sec/1_intro}
\input{sec/2_related_works}
\input{sec/3_methods}
\input{sec/4_security}
\input{sec/5_exp}
\input{sec/6_conclusion}

\section*{Acknowledgment}

The authors would like to thank the reviewers for their meticulous evaluation of this work and their instructive comments.




\bibliographystyle{IEEEtranS}
\bibliography{ref}
%



\end{document}

%% file: sec/1_intro.tex
\section{Introduction}
Due to the increasing demand for security in daily life, more and more applications are adopting biometric technology for identification and authentication. Biometric technologies store unprotected biometric templates directly in the database and use them to authenticate users. However, these unprotected biometric templates can lead to serious security issues. For example, if an attacker manages to steal unprotected biometric templates stored in the database through a system vulnerability, they could use these templates to reconstruct the original biometric data. Since each user’s biometric data are immutable, once the unprotected biometric templates stored in a database are compromised, the corresponding application registered with this biometric data will be invalidated. Therefore, privacy and security issues are very important in biometric templates, and BTP methods are proposed. The international standard ISO/IEC 24745\cite{ref57} states that BTP method should satisfy irreversibility, revocability and unlinkability. Irreversibility requires that it should be difficult to reconstruct the original biometric data from the corresponding protected biometric template. Revocability requires that when a protected biometric template is compromised, it should be easy to generate a new protected biometric template. Unlinkability requires that different templates from different applications cannot be used for cross-matching.
\begin{figure*}
    \centering
    \includegraphics[width=0.9\linewidth]{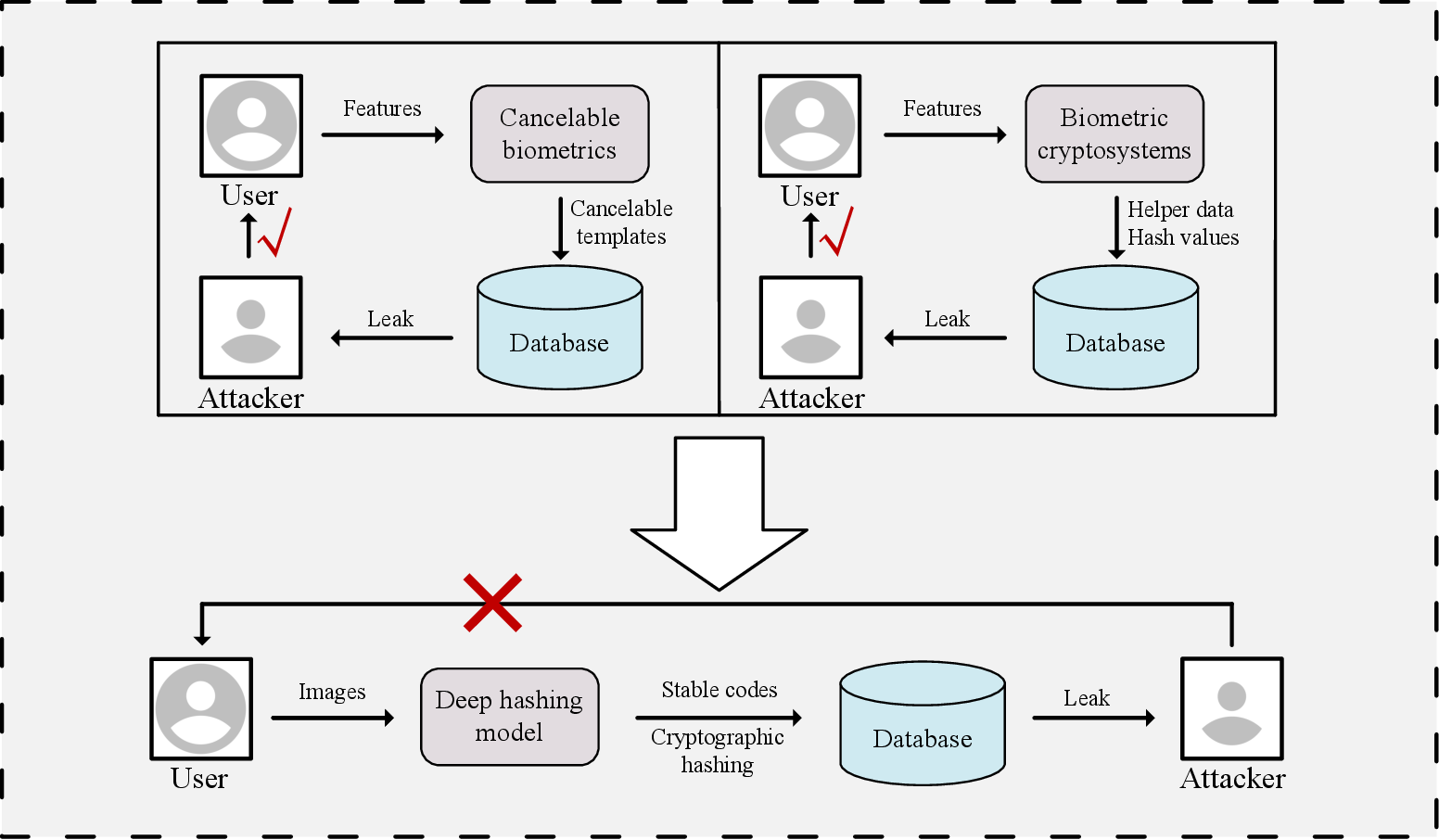}
    \caption{Research Motivation}
    \label{fig:enter-label}
\end{figure*}

In order to satisfy the above three properties, many BTP methods have been proposed, and the existing BTP methods can be categorized into: (1) cancellable biometrics (CB) \cite{ref5,ref6,ref9,ref7,ref12,ref15}, (2) biometric cryptosystems (BCS) \cite{ref17,ref18,ref19,ref20,ref21,ref22}, (3) methods using NN to learn pre-defined templates \cite{ref34,ref35}, (4) methods using NN to learn latent templates \cite{ref32,ref33}.

CB methods irreversibly transform the original biometric data, use different parameters to obtain different protected biometric templates for the same biometric data, and match the protected biometric templates by comparing the similarity between them with a threshold value. However, the irreversibility of current CB methods is difficult to quantify, as the transformed protected templates still contain some information from the original biometric data \cite{ref1}. In addition, numerous CB methods have been demonstrated to be vulnerable to various attacks \cite{ref10,ref11,ref8,ref13,ref14,ref16}, which suggests that these methods are unable to fulfill the requirements of irreversibility and unlinkability. In BCS, the original biometric data is combined with an encryption key to form helper data, which is stored in a database. Users recover the encryption key in conjunction with helper data, and verify the validity of the key to indirectly achieve matching. Unlike CB methods, BCS use cryptographic hashing to ensure strong security of stored data. Due to the use of cryptographic hashing, BCS has to introduce ECC to overcome intra-class variations in biometric data and to achieve fuzzy recognition that tolerates certain errors. However, numerous works \cite{ref23,ref24,ref25,ref26,ref27,ref28,ref51} have shown that the introduction of ECC, along with the storage of helper data, can cause serious privacy issues for BCS. In methods using NN to learn latent templates, representations learned by these methods can only map similar images to similar codes instead of a stable code. In these methods, the mapped code is usually combined with CB methods or BCS methods for security \cite{ref34,ref35}. However, as we discussed previously, both the CB methods and the BCS methods have some privacy issues. Therefore, the security of methods using NN to learn latent templates cannot be guaranteed either.

Among the above three kinds of methods, the protected biometric templates in the CB methods as well as methods using NN to learn latent templates still contain some information of the original biometric data, so it is difficult for these methods to guarantee strong security. Whereas, the use of cryptographic hashing in the BCS scheme successfully guarantees strong security (without considering the privacy issues introduced by ECC). Through the above analysis, we believe that a strongly secure BTP method that ultimately stores templates should use cryptographic level processing like cryptographic hashing to guarantee strong security. But it is a fact that the use of cryptographic hashing seems infeasible due to the intra-class variability of biometric data. As shown in Fig. 1, we get the idea of using NN to map different biometric data of the same user to a stable code to address the above issues. Through this idea, we can store only the data processed using cryptographic hashing, avoid introducing ECC that leads to additional privacy risks, and achieve strong security.

In fact, similar work \cite{ref32,ref33} has been proposed in the methods using NN to learn pre-defined templates. However these methods still have problems with security and revocability. Methods using NN to learn pre-defined templates predefine a user-specific random binary code. Then they learn the mapping from the user's biometric data to that code through a NN. In this way mapping biometric data of the same user to a stable code can be achieved. However, the revocability of these methods is difficult to guarantee due to the use of pre-defined codes. Specifically, each time a user is revoked and regenerated, the pre-defined code corresponding to the user needs to be replaced and the entire NN needs to be retrained. This implementation of revocability is clearly unrealistic. In addition, in these methods it is necessary to store training samples (unprotected biometric data) in order to retrain the model, which also poses a serious privacy risk.

Based on the previous description, we can find that existing BTP methods cannot guarantee both strong security (for CB, BCS, NN learns latent templates) and reasonable revocability (for NN learns pre-defined templates). The property that deep hashing can map similar images to similar codes and even ensure that the distance between similar codes is below a given hamming radius threshold inspires us to utilize deep hashing to map similar images to stable codes, thus we propose a framework called BioDeepHash based on deep hashing and cryptographic hashing to solve the previous problems. The framework employs a deep hashing model to eliminate the noise among the biometric data. By integrating three loss functions, the model can map similar biometric data of the same user to a stable code. Throughout the biometric identification process, only the cryptographic hash values are stored. The security of these cryptographic hash values has been verified. To achieve revocability, we introduce a publicly accessible, application-specific XOR string that is independent of the biometric data. By modifying the XOR string, we can easily revoke and regenerate the stored templates. The framework has been tested on facial and iris data and obtained excellent performance. The contributions of this paper are summarized as follow:

\begin{itemize}
    \item We propose a BTP framework named BioDeepHash that employs deep hashing and cryptographic hashing to satisfy irreversibility, revocability and unlinkability. In our framework, there is no need to store any helper data related to biometric data, which significantly enhances the security of our framework.
    \item We design the deep fuzzy hashing (DFH) model for the field of BTP. This model combines class-wise loss function, class label regression loss and quantization loss to achieve the task of mapping biometric data of same users to stable codes. Depending on this model, we are able to effectively identify unregistered users and achieve an extremely low FAR.
    \item We employ an application-specific XOR string to achieve revocability within our framework. Furthermore, application-specific XOR string is also used as salt value to salt the output of the DFH model, preventing rainbow table attack.
    \item We conduct extensive experiments on iris and facial datasets to validate the performance of our framework. Compared to existing BTP methods, our method demonstrates superior performance. Additionally, we assess the security and privacy of our framework based on various attack methods.
\end{itemize}

%% file: sec/2_related_works.tex
\section{Related Works}
In this section, we will introduce existing BTP methods. We roughly classify BTP methods into two categories: traditional BTP methods and NN-based BTP methods.
\subsection{Traditional BTP Methods}
Traditional BTP methods can generally be categorized into two types: CB \cite{ref5,ref6,ref9,ref7,ref12,ref15} and BCS \cite{ref17,ref18,ref19,ref20,ref21,ref22}. 
\subsubsection{Cancelable Biometrics}
The concept of CB was first introduced by Ratha \cite{ref5}, primarily based on irreversible transformations. Subsequently, many methods of CB have been proposed. Teoh et al. \cite{ref6} proposed a method for iris template protection that based on random projection. In their research, they employed random projection matrices to transform original biometric data into another space, thereby generating cancelable protected templates. Rathgeb et al. \cite{ref9} proposed a CB method based on Bloom filters. In their method, Bloom filters are used to perform irreversible transformations on biometric data, generating cancelable protected templates. However, \cite{ref10,ref11} has demonstrated that BTP methods based on Bloom filters fail to meet the unlinkability. Subsequently, Zhao et al. \cite{ref7} proposed a CB method based on local ranking,  they used ranking values to generate cancelable protected templates. However, Ouda presented a scheme in \cite{ref8} that demonstrated issues with the irreversibility of Zhao's method. Jin et al. \cite{ref12} proposed a method based on Index-of-Max hashing (IoM). Inspired by Locality-Sensitive Hashing, this method transformed biometric features into discrete indices (maximum ranking) to generate hashed codes, thereby creating cancelable protected templates. Subsequently, an attack method named as the known sample attack was proposed \cite{ref13}, demonstrating that most BTP methods that utilize distance-preserving hashing are vulnerable to this type of attack. Additionally, another method named as the similarity attack \cite{ref14} demonstrated that methods based on distance-preserving properties could lead to privacy breaches. This was further evidenced through experiments that showed the vulnerability of Bloom filter based methods under such attacks. In recent years, to address the issue of pre-alignment of biometric features, Lee et al. \cite{ref15} proposed a BTP method based on Random Augmented Histogram of Gradients (R·HoG). This method utilized column vector random augmention technique combined with orientation gradient histogram to generate alignment-robust cancelable templates. However, it was demonstrated in \cite{ref16} that the R·HoG method can not satisfy unlinkability.
\subsubsection{Biometric Cryptosystems}
BCS essentially perform matching of biometric feature templates within an encrypted domain. The principal schemes include the fuzzy vault \cite{ref17,ref18}, fuzzy commitment \cite{ref19,ref20}, and fuzzy extractor \cite{ref21,ref22}. Specifically, the fuzzy vault scheme secured a key using a set of unordered points, creating a vault that required polynomial reconstruction techniques to retrieve the key. However, it has been indicated in \cite{ref23,ref24,ref25} that the fuzzy vault scheme might leak original biometric data and was vulnerable to cross-matching attack. The fuzzy commitment scheme combined ECC with cryptographic techniques, creating a commitment by associating biometric data with a random codeword. The codeword was then recovered using error-correcting techniques for verification. However, it has been demonstrated in \cite{ref26,ref27} that the fuzzy commitment scheme was vulnerable to decodability attack. Subsequently, \cite{ref28} proposed an improved scheme to prevent cross-matching. Rathgeb et al. \cite{ref51} conducted a statistical attack on the fuzzy commitment scheme, successfully retrieving the stored key, which indicated that the fuzzy commitment scheme still posed risks of privacy leakage. The fuzzy extractor scheme  extracted consistent random strings and helper data from biometric data. During the reconstruction process, the helper data was used with similar biometric data to reconstruct the consistent random string. In this scheme, the extracted random string was used as the encryption key. However, \cite{ref29} pointed out that the fuzzy extractor scheme struggled to ensure  unlinkability and irreversibility when multiple sketches were exposed, leading to potential privacy breaches. From the discussion above, we can find that most existing traditional BTP methods fail to simultaneously satisfy irreversibility and unlinkability.

\subsection{NN-based BTP Methods}
In recent years, NN has been widely applied in the field of biometrics due to its strong capability in feature extraction from images. Consequently, an increasing number of NN-based BTP methods have been proposed. We categorize NN-based BTP methods into two types with reference to the classification in \cite{ref4}: Non-NN and NN-learned. Since Non-NN methods only use NN for feature extraction, essentially the methods to achieve the protection effect still use traditional BTP methods, so we focus on NN-learned methods. The NN-learned methods refer to using NN to learn a transformation method that directly converts biometric data into protected templates. And NN-learned methods can be divided into two categories: method using NN to learn latent templates \cite{ref34,ref35} and method using NN to learn pre-defined templates \cite{ref32,ref33}.

\subsubsection{The method using NN to learn latent templates}
One implementation of the NN-learned methods is to use a NN to learn the representation of the protected template itself. In \cite{ref34}, the extracted feature vectors were arranged using a specific matrix, and then a NN was employed to simulate the IoM algorithm to learn the mapping from the feature vectors to the protected templates. The essence of this method remained a CB method, and the final matching still utilized a similarity matching method. In \cite{ref35}, a random CNN with a random triplet loss was proposed. This CNN transformed biometric data along with user-specific keys into protected templates, and employed the method of secure sketches instead of storing keys. Due to the use of secure sketches which incorporated ECC, this approach might also be susceptible to decodable attacks and statistical attacks. Since the representations learned may exhibit some intra-class variability (i.e. differences between the same user's enrollment and various authentication attempts), the aforementioned methods do not apply cryptographic hashing to process the protected templates and thus could not provide strong security guarantee. 

\subsubsection{The method using NN to learn pre-defined templates}
Another implementation of the NN-learned methods is to predefine a user-specific random binary code as the label of biometric data and then learn a mapping from the user's biometric data to that code via a NN. These methods are primarily based on the methods proposed in \cite{ref32}. In \cite{ref32}, each user was assigned a random maximum entropy binary code, and a CNN was trained to map the user's biometric data to this binary code. Finally, cryptographic hashing function was applied to the binary code to generate the final protected template. In \cite{ref33}, the previously used maximum entropy binary codes were replaced with Low-Density Parity-Check (LDPC) codes. By leveraging the error-correcting capabilities of LDPC codes, the scheme enhanced robustness to intra-class variations. Although the predefined binary code scheme achieves high security through cryptographic hashing function, it faces significant challenges in terms of revocability. When a user's template is lost and needs to be regenerated, the entire network requires retraining. In addition, in order to retrain the model these methods need to store unprotected training samples, which can lead to serious privacy risks.

%% file: sec/3_methods.tex
\section{BioDeepHash}
\begin{figure*}
    \centering
    \includegraphics[width=\textwidth]{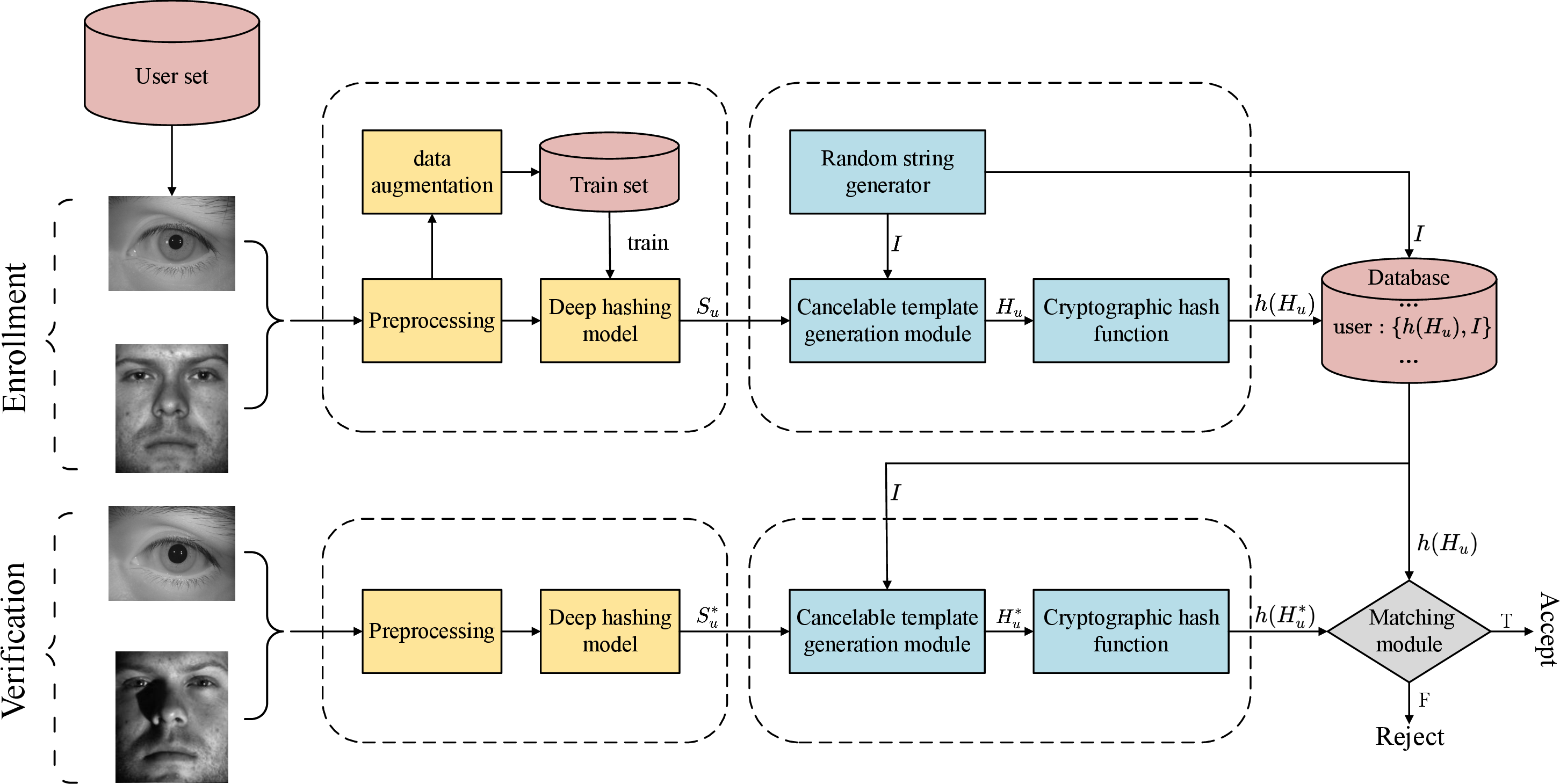}
    \caption{An illustration of the proposed framework}
    \label{fig:enter-label}
\end{figure*}

According to our previous analysis, a strongly secure BTP method requires the use of cryptographic hashing, and considering the avalanche effect of cryptographic hashing, error correction before cryptographic hashing is an important task. We define the error correction task as the following problem:
Given a set of samples $X=\{x_{1},x_{2},\cdots ,x_{n-1},x_{n}\}$, label set $Y =\{y_{1} ,y_{2},\dots ,y_{n-1},y_{n}\}$, class center set 
$\mathcal{C}  =\{c_{1},c_{2},\dots ,c_{m-1},c_{m}\}$, where $n$ is the number of samples, and $m$ is the number of classes. We assume that each of the aforementioned samples is generated by adding a certain amount of noise to its corresponding class center. 
\begin{equation}
    x_{i} =c_{y_{i}}  + \varepsilon_{i}
\end{equation}
where $\varepsilon_{i}$ denotes the error term for each sample caused by factors such as lighting, pose, and rotation. We need to find a mapping function \( f \) such that $f(x_{i}) = f(x_{j})$, if $y_{i}=y_{j}$, otherwise, $f(x_{i}) \ne f(x_{j})$. Although many schemes in BCS use ECC to accomplish similar tasks, numerous attacks targeting ECC have been proposed in \cite{ref26,ref27}. In our framework, we utilize a deep hashing model to learn the mapping from biometric data of the same user to a stable code for the task of error correction. The specific details of deep hashing model will be described in Section \ref{sec:method:dfh}. Next, we consider our framework's revocability. A straightforward approach to achieving revocability using deep hashing is to retrain entire NN, but this significantly increases time costs. Moreover, a prerequisite for retraining is the storage of training samples (unprotected biometric data) of the NN, which can lead to serious privacy risks. To enhance efficiency and ensure that the framework does not introduce additional privacy risks, we propose a framework based on deep hashing and application-specific XOR string. This framework ensures both security and high recognition accuracy.

\input{sec/algo1}

\subsection{System Overview}
The overall scheme of the proposed framework is illustrated in Fig. \ref{fig:enter-label}. The flow of the enrollment phase is shown in Algorithm \ref{algo:algo1}, data is collected from users who register. These data are augmented by data augmentation and preprocessing to generate a training set $\mathcal{T}$ to meet the data requirements for training. Using these processed data, a deep hashing model $f(\Theta;x)$ is trained to map the biometric data \( x_u \) of the user to a stable code. Subsequently, \( x_u \) is input into model $f(\Theta;x)$, which generates the stable code \( S_u \) for each user with sgn function. An application-specific string \( I \) is assigned to each user, and the cancelable template generation module is utilized to convert \( I \) and \( S_u \) into the cancelable template \( H_u \). \( H_u \) is then processed using the cryptographic hash function to produce the secure hash code \( h(H_u) \). Finally, both \( h(H_u) \) and \( I \) are stored in the database. Additionally, the application-specific string \( I \) introduced in the registration phase is not a security parameter, meaning that this data can be exposed to any user without posing any privacy risk. A detailed analysis of this will be provided in Section \ref{sec:method:gene}. 

In the verification phase, the biometric data $x_u^*$ of the user is input into model $f(\Theta;x)$, yielding the corresponding stable code $ S_u^*$ with sgn function. The code $S_u^*$, along with the user-specific XOR string $I$, is processed through the cancelable template generation module to produce cancelable template $H_u^*$. $H_u^*$ is then processed using a hash function to obtain the secure hash code $h(H_u^*)$. Finally, $h(H_u^*)$ enters the matching module, and there are two matching strategies in this module, the details of which will be described in 3.4. When the matching module returns 1, user authentication is successful. Otherwise, user authentication is rejected. In the other parts of this section, we will introduce the specific modules within the framework.

\subsection{Deep Fuzzy Hashing}\label{sec:method:dfh}

The deep hashing model is primarily used to learn a mapping function that maps images within the same class to hash codes that are as similar as possible. Given a set of samples $X$ and a corresponding set of labels $Y$, the task of the deep hashing model is to learn a mapping function $f(\Theta;x)$, where $\Theta$ is a set of network parameters. In our framework, to ensure security, we apply cryptographic hashing to the protected templates obtained. Considering the avalanche effect inherent in cryptographic hashing, where small changes in the input lead to significant changes in the output, it is necessary to perform error correction on the inputs to the cryptographic hashing. In our framework, we design DFH model to perform the task of error correction. The DFH used in our framework imposes stricter constraints compared to typical deep hashing models. Specifically, images from the same user are mapped to codes that are not just similar but completely stable through our DFH.

In our framework, we design the DFH model, which combines deep class-wise hashing (DCWH) loss \cite{ref38} with class label regression loss \cite{ref39}. This integration aims to map images of the same user to a stable code. Next, we will provide a detailed introduction to the DFH, divided into the following two sections.
\subsubsection{Model Architecture}
The DFH we use primarily consists of feature extraction and hashing layers. The feature extraction layer is mainly composed of convolutional layers, pooling layers, and fully connected layers. In order to extract more discriminative biometric features, we design a simple convolutional neural network (CNN), and the specific structure of which is illustrated in the Fig. 3 below. After extracting features from the biometric data through the feature extraction layer, these features are then fed into a hashing layer composed of fully connected layers, which transforms the extracted features into hash codes of a specified length $L$. Finally, the output from the hashing layer is processed using the sign function, generating binary codes that serve as the stable code in our framework.
\begin{figure}[h]
  \label{fig:fig3}
  \centering
  \includegraphics[width=\linewidth]{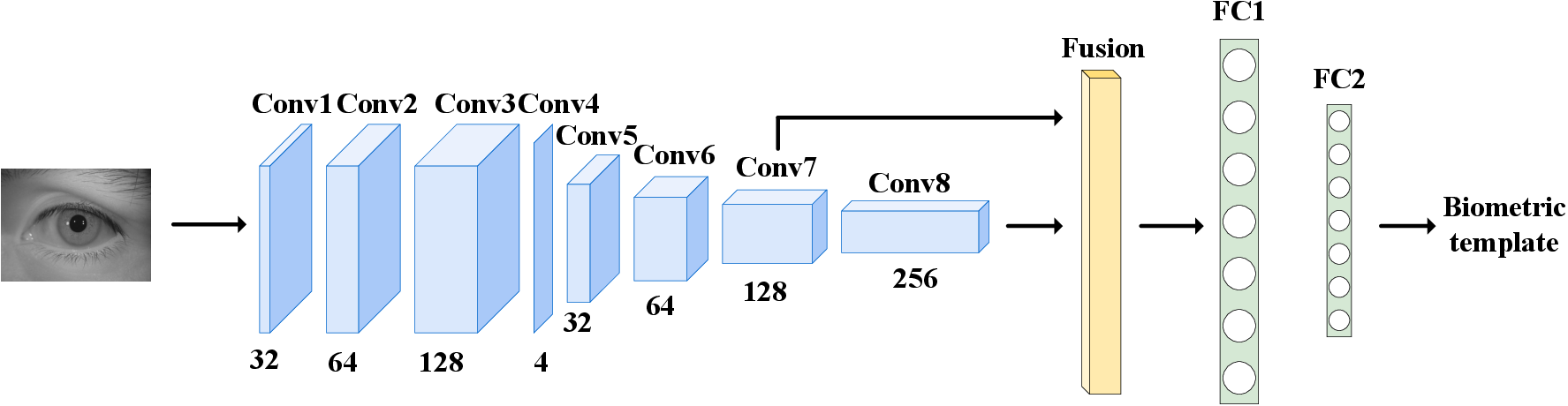}
  \caption{Model Architecture}
\end{figure}
\subsubsection{Loss Function}
In DFH, the loss function is composed of three main parts:
\paragraph{Class-wise loss function based on a Gaussian model}
The loss is derived from the DCWH approach \cite{ref38}. The main idea behind this loss is to better utilize the information within class labels. The DCWH loss, which is derived from the pair-wise loss \cite{ref48}, is formulated as follows:
\begin{multline}
    \min_{\Theta}\sum_{i=1}^{n} \{(1-P_{i,j})  d(f(\Theta;x_{i} ),f(\Theta;x_{j} ))\\
     -P_{i,j}d(f(\Theta;x_{i} ),f(\Theta;x_{j} ))\}
\end{multline}
where \(d(\cdot)\) represents the distance function, \(x_{j}\) is an anchor image and $P_{i,j}$ is a pair-wise label which can be calculated by the following formula:
\begin{equation}
    P_{i,j} =\left\{\begin{matrix} 1 & y_{i}\ne y_{j} \\0&y_{i}= y_{j} \end{matrix}\right. 
\end{equation}
The main idea of the pair-wise method is to reduce the distance between the binary codes of similar instances and increase the distance between dissimilar instances. Following this, the triplet loss \cite{ref49} and quintuplet loss \cite{ref50} were introduced. Inspired by these methods, DCWH utilizes the distances of multiple sample pairs in the loss function calculation to enhance performance. Let us consider such an extreme situation where the loss is measured by calculating the distance between each sample and all other samples, and calculation formula is as follows:
\begin{multline}
    \min_{\Theta} \sum_{i=1}^{n} \{ \sum_{j=1}^{n_{is}} d(f(\Theta; x_i), f(\Theta; x_j)) \\
     - \sum_{k=1}^{n_{id}} d(f(\Theta; x_i), f(\Theta; x_k))  \}
\end{multline}
where $n_{is}$ is the number of samples with the same class label as $x_{i}$, and $n_{id}$ is the number of samples with a different class label from $x_{i}$. However, in CNN, only a small number of images can be trained in a batch, thus making the above calculation method impractical. Therefore, in DCWH, a compromise approach is adopted, where the loss function for each sample is calculated based on the difference between it and the centers of all classes, instead of the difference with all other samples and calculation formula is as follows:
\begin{multline}\label{eq:eq5}
    \min_{\Theta} \sum_{i=1}^{n} \{   d(f(\Theta; x_i), f(\Theta; c_{y_i})) \\
     - \sum_{c_{y_j} \in C^*}  d(f(\Theta; x_i), f(\Theta; c_{y_j})) \}
\end{multline}
where $C^*$ is a set of class labels excluding \( y_i \). To enable deep hashing model to learn a more compact space, DCWH employs a gaussian distribution-based normalized probability model to improve formula \ref{eq:eq5}:
\begin{multline}
    \label{eq:eq6}
    P(y_{i} |x_{i} ; \Theta)=\\
    \frac{\frac{1}{\sqrt{2\pi}\sigma_{y_{i}}}\mathrm{exp} \{-\frac{1}{2\sigma^{2} _{y_{i}} }d^{2}(f(\Theta; x_i),f(\Theta; c_{y_i}))  \}}{ {\textstyle \sum_{j=1}^{|C|}}\frac{1}{\sqrt{2\pi}\sigma_{y_{j}}}\mathrm{exp} \{-\frac{1}{2\sigma^{2} _{y_{j}} }d^{2}(f(\Theta; x_i),f(\Theta; c_{y_j}))  \}  } 
\end{multline}
where $\sigma^{2} _{y_{i}}$ represents the variance of the class center \( y_i \) corresponding to \( x_i \). In DFH model, the set of class centers $\mathcal{C}$ is treated as a learnable parameter and is updated during the training process. Finally, by applying the negative log-likelihood to formula \ref{eq:eq6}, we will derive the  formula of class-wise loss function $L_{1}$.
\begin{multline}\label{eq:eq7}
    {\min_{\Theta,\mathcal{C}}}\text{ }L_{1} = \\
    -\sum_{i=1}^{n} \mathrm{log} \frac{\frac{1}{\sqrt{2\pi}\sigma_{y_{i}}}\mathrm{exp} \{-\frac{1}{2\sigma^{2} _{y_{i}} }d^{2}(f(\Theta; x_i),f(\Theta; c_{y_i}))  \}}{ {\textstyle \sum_{j=1}^{|C|}}\frac{1}{\sqrt{2\pi}\sigma_{y_{j}}}\mathrm{exp} \{-\frac{1}{2\sigma^{2} _{y_{j}} }d^{2}(f(\Theta; x_i),f(\Theta; c_{y_j}))  \}  } 
\end{multline}
\paragraph{Regression loss and quantization loss}
To enable our DFH model to map images of the same class to identical binary strings, we adopted the label regression matrix proposed in FSDH \cite{ref39}. By using such a regression matrix, labels are linked to binary codes, facilitating the generation of identical hash codes for different samples from the same user. The formula for regression loss is as follows:
\begin{equation}
    {\min_{\mathbb{T},W}}\text{ } L_{2}  =\sum_{i=1}^{N}||T_{i}-W^{T}y_{i}|| ^{2} 
\end{equation}
where $\mathbb{T}=\{T_{1},T_{2},\dots,T_{n-1},T_{n}\}$, $T_{i}$ is the output of the $i$-th user through the deep hashing model and \( W \) is the regression matrix. Then the loss function is updated to formula 9.
\begin{equation}
    {\min_{\Theta,\mathcal{C},\mathbb{S},W}}\text{ } L_{1}+L_{2}
\end{equation}
In our approach, to obtain the final binary codes, we use the sign function to convert \( T_i \) into \( S_i \), introducing a discrete optimization problem. Consequently, Formula 9 can be approximated as follows:
\begin{equation}
    \begin{aligned}
    & {\min_{\Theta,\mathcal{C},\mathbb{S},W}}\text{ } L_{1}+L_{2}\\
    & s.t. \text{ } S_{i}=sgn(T_{i})
\end{aligned}
\end{equation}
We adopt the approach in \cite{ref40} to address the aforementioned discrete optimization problem, introducing a quantization loss between \( S_i \) and \( T_i \). The updated loss function is as follows:
\begin{equation}\label{eq:eq11}
    {\min_{\Theta,\mathcal{C},\mathbb{S},W}}\text{ } L_{1}+L_{2}+\eta\sum_{i=1}^{n} ||S_{i}-T_{i}||^{2} _{2} 
\end{equation}
where \(\eta\) is the Lagrange multiplier. We use Formula \ref{eq:eq11} as the final loss function for DFH model.
\input{sec/algo2}
\subsection{Cancelable Template Generation Module}\label{sec:method:gene}
In our framework, we use the DFH model mentioned in the previous section to map biometric data of the same user to a stable code. However, this method does not satisfy revocability without retraining the model. We propose a secure and efficient method by introducing an application-specific XOR string $I$ to achieve revocability in the BioDeepHash. Additionally, a portion of the XOR string is used as a salt combined with the template to defend against rainbow table attack \cite{ref52}. Specifically, when a user's data is processed through the DFH model to produce a code \( S_u \) of length \( L \), we generate a random XOR string of length \( 2L \) for each user. Subsequently, we use the following formula to obtain the cancelable template:
\begin{equation}
    H_{u} =(S_{u}\oplus I_{1})\text{ }||\text{ }I_{2} 
\end{equation}
where $I_{1}$ represents the first \(L\) bits of $I$, and $I_{2}$ represents the last \(L\) bits of $I$. When a user needs to revoke and regenerate, he can simply regenerate the corresponding XOR string. 

Assuming that the protected template stored in the database by a user is lost, an attacker obtains \( h(H_{u}) \), where \( H_{u} = (S_{u}\oplus I_{1})\text{ }||\text{ }I_{2}  \), when we replace it with the XOR string \( I^{*} \), the new template \( h(H_{u}^{*}) \), where \( H_{u}^{*} = (S_{u}\oplus I_{1}^{*})\text{ }||\text{ }I_{2}^{*}  \), will be completely different due to the properties of cryptographic hashing. Furthermore, the introduction of the XOR string does not pose additional privacy risks. Consider a situation where an attacker possesses the XOR string \( I \) and \( h(H_{u}) \), where $H_{u} =(S_{u}\oplus I_{1})\text{ }||\text{ }I_{2}$. If the attacker wishes to obtain $S_{u}$ using $I$ and $h(H_{u})$, he would need to invert \( h(H_{u}) \). However, due to the properties of cryptographic hashing, this is unachievable. Therefore, attackers are unable to reverse any information related to the original biometric data from these data.
\subsection{Cryptographic Hash and Matching Strategy}
\subsubsection{Hash Function}
In the above steps, we have addressed the issues of error correction and revocability. Finally, to ensure that the data stored in the database do not leak any information about the original biometric data, we use a cryptographic hash function to process and obtain the final protected template. In our framework, we use SHA3-512 to generate the protected template. SHA3-512, as a type of hash function, exhibits a substantial avalanche effect. Specifically, when a single bit in the input is changed, each bit in the output has a $50\%$ chance of undergoing a change. This property provides an important safeguard for the security of our framework.
\subsubsection{Matching Strategy}
In our framework, we use two different matching strategies considering the different requirements of the verification phase. The flow of the two matching strategies is shown in Algorithm \ref{algo:algo2}. The first matching strategy adopts a precise matching approach, validation data $x_{u}$ are sequentially fed into the DFH model, cancelable template generation module to obtain the $H_{u}$, and then processed using cryptographic hashing to obtain the secure hash code $h(H_{u})$. Ultimately, $h(H_{u})$ is used to match against protected templates in the database in turn, and user authentication is only accepted if $h(H_{u})$ is identical to a protected template in the database. Another matching strategy we refer to the approach in \cite{ref33}. In the validation process, we first crop the validation data $x_{u}$, specifically, determine a cropping length $k$. Assuming that the image size is \(a \times b\), crop out all the images with size \((a-k) \times (b-k)\), resulting in \((k+1) \times (k+1)\) images. Then, each image is processed using the first strategy. Subsequently, we calculate the matching score: \( count/(k+1)^2 \), where \( count \) is the number of successfully matched images. We can set a threshold $\tau$, if the matching score exceeds $\tau$, the user's verification is accepted, otherwise the user's verification is denied.

%% file: sec/algo1.tex
\begin{algorithm}
\label{algo:algo1}
\caption{Enrollment process}\label{algo:algo1}
\KwIn{sample set $X$, label set $Y$, DFH model $f(\Theta;x)$, Application-specific XOR string $I$, SHA3-512 function $h()$}
\KwOut{model parameter $\Theta$, template database $DB$}
$X$ $\leftarrow$ prepossess($X$)\;
$L$ $\leftarrow$ sizeof($X$)\;
$\mathcal{T}$ $\leftarrow$ $\emptyset$\;
\For{$i=1$  $\mathrm{to}$  $L$}{
    $\mathcal{T}$ = $\mathcal{T}$ $\bigcup$ $\{x_{i},y_{i}\}$\;
}
train model $f(\Theta;x)$ using $\mathcal{T}$ and keep updating $\Theta$\;
\For{$i=1$  $\mathrm{to}$  $L$}{
    $T_{i}$ $\leftarrow$ $f(\Theta;x_{i})$\;
    $S_{i}$ $\leftarrow$ sgn($T_{i}$)\;
    $H_{i}$ $\leftarrow$ cancelable\_Template\_Generate($S_{i},I$)\;
    $DB$ $\leftarrow$ $DB$ $\bigcup$ $h(H_{i})$\;
}
\Return {$\Theta$, $DB$}\;
\end{algorithm}

%% file: sec/algo2.tex
\begin{algorithm}
\label{algo:algo2}
\caption{Matching strategy in the verification phase}\label{algo:algo2}
\KwIn{Validation data $x_{u}$, DFH model $f(\Theta;x)$, Application-specific XOR string $I$, cropping length $k$, matching strategy $\mathbb{M}$, template database $DB$, threshold $\tau$}
\KwOut{0/1}
\eIf{$\mathbb{M}$ is the precise matching strategy}
{
    $S_{u}$ $\leftarrow$ $f(\Theta;x_{u})$\;
    $H_{u}$ $\leftarrow$ cancelable$\_$Template$\_$Generate($S_{i},I$)\;
    \If{$h(H_{u})$ $\in$  $ DB$}{
        \Return $1$\;
    }
}
{
    $count$ $\leftarrow$ 0\;
    $\mathcal{X}$ $\leftarrow$ crop($x_{u}$,$k$)\;
    $num$ $\leftarrow$ \((k+1) \times (k+1)\)\;
    \For{$i=1$  $\mathrm{to}$ $num$}{
    $S_{i}$ $\leftarrow$ $f(\Theta;\mathcal{X}_{i})$\;
    $H_{i}$ $\leftarrow$ cancelable\_Template\_Generate($S_{i},I$)\;
    \If{$h(H_{i})$ $\in$ $ DB$}{
        $count$ $\leftarrow$ $count$ $+$ $1$\;
        }
    }
    $matching$ $score$ $\leftarrow$ $count $ / $num$\;
    \If{$matching$ $score$ $\ge$ $\tau$}{
        \Return 1\;
    }
}
\Return 0\;
\end{algorithm}

%% file: sec/4_security.tex
\section{Security Analysis}

\subsection{Threat Model}
In order to analyze the security of BioDeepHash, we must first define the threat model. In this paper, we use the most difficult threat model specified by ISO/IEC 30316 which is called the full disclosure model. This model assumes that the attacker knows all information about the protection method (e.g. algorithms, secrets, model, etc.). Since this type of attacker represents the worst case scenario in practice, we decided to analyze the security of BioDeepHash based on the full disclosure model. Next we present the definition of the full disclosure model in our method.
\begin{itemize}
    \item \textbf{Attacker's goal}: The attacker aims to impersonate a user enrolled in the BioDeepHash framework.
    \item \textbf{Attacker's knowledge}: 
    \begin{enumerate}
    \item Matching strategy  $\mathbb{M}$
    \item Application-specific XOR string \( I \)
    \item Protected templates for users stored in the database $h(H_{u})$
    \item White-box knowledge of the DFH model $f(\Theta;x)$ (including model structure and model parameters $\Theta$)
    \end{enumerate}
\end{itemize}
It is worth mentioning that the attacker cannot directly obtain the confidence vector  and the output of the DFH model corresponding to the enrolled user because they are discarded after the enrollment phase.

\subsection{Irreversibility Analysis}\label{sec:irre}
In BioDeepHash, we only store the protected templates encrypted using cryptographic hash functions and application-specific XOR strings in the database. As mentioned in Section \ref{sec:method:gene}, the application-specific XOR strings introduced by our method do not leak any biometric data-related information even if lost. For an attacker, even if they manage to steal the protected templates stored in the database, they would still need to invert the cryptographic hash function, which is computationally infeasible. 

Under the scenario where cryptographic hash function is irreversible, we consider the possibility of brute-force (BF) attack. Based on the discussion in Section \ref{sec:bf}, the most advantageous strategy for a BF attacker would be to target the output of the DFH model. This means that the irreversibility of BioDeepHash primarily depends on the length \( L \) of the DFH model output. Therefore, choosing an appropriately \( L \) is crucial for ensuring the security of BioDeepHash against BF attack. In our method, we set the value of \( L \) to be 96, 120, and 144. Additionally, we can adjust \( L \) based on specific security needs to tailor the level of security provided.

In addition to the above attack points, attackers can also focus their attacks on DFH model, similar to the template inversion (TI) attack \cite{ref60}, which utilizes deep hashing models to invert stored templates to get the original biometric data. However, the protected templates stored in our framework are processed using SHA512 encryption, and attackers are unable to use these protected template for effective inversion attacks. Based on this scenario, an attacker who wants to successfully perform a TI attack needs to obtain the output of the training data in the DFH model, which is not directly accessible in BioDeepHash. Therefore, we can define the TI problem against the BioDeepHash as obtaining the output of the training data in the DFH model or the training data itself, given the DFH model $f(\Theta;x)$, the model parameters $\Theta$, and any auxiliary information available to the attacker. In Section \ref{sec:TIA} we analyze a kind of TI attack \cite{ref61} that may be applicable to such scenarios, proving that BioDeepHash can defend against such attack.

\subsection{Revocability Analysis}\label{esc:security:revo}
Revocability refers to the ability of a BTP method to ensure that if a protected template is lost or compromised, it can be revoked and a new template can be generated. Revocability in BioDeepHash relies on the cancelable template generation module. Specifically, when the protected template $h(H_{u1})$ is lost and needs to be regenerated, we simply replace the corresponding application's XOR string $I$, regenerate the new cancellable template $H_{u2}$ by combining the outputs of $S_{u}$ in the DFH model with the new XOR string $I^*$ via the cancellable template generation module, and use cryptographic hash processing to get the new protected template $h(H_{u2})$. Through the above method, BioDeepHash can achieve revocability without the need to retrain the model.
\subsection{Unlinkability Analysis}
Unlinkability refers to attackers unable to determine whether two templates from different applications correspond to the same user. We employ the framework proposed in \cite{ref41} to evaluate the unlinkability of the protected templates generated by BioDeepHash. Specifically, the framework analyzes unlinkability by examining the overlap between the score distributions of mated and unmated samples. Mated samples refer to protected templates generated from the biometric data of the same user registered in different applications, while unmated samples refer to protected templates generated from the biometric data of different users registered in different applications. Based on this concept, the framework provides two different metrics for assessing unlinkability: 1) Local measure $D_{ \leftrightarrow} (s)$.
This metric reflects the unlinkability at a specific distance score $s$. 2) Global measure $D_{ \leftrightarrow}^{sys}$. This metric represents the overall unlinkability of the system. Both metrics range from [0,1], where values closer to 0 indicate better unlinkability. In our experiments, we assign two different XOR strings for two applications, generating the corresponding protected templates to simulate user enrollment under two different applications. The experimental results for assessing unlinkability are shown in Fig. 4, indicating low $D_{ \leftrightarrow} (s)$ across most scores. The $D_{ \leftrightarrow}^{sys}$ is 0.0200, close to 0. This demonstrates that BioDeepHash meets the unlinkability.
\begin{figure}[h]
  \centering
  \includegraphics[width=.9\linewidth]{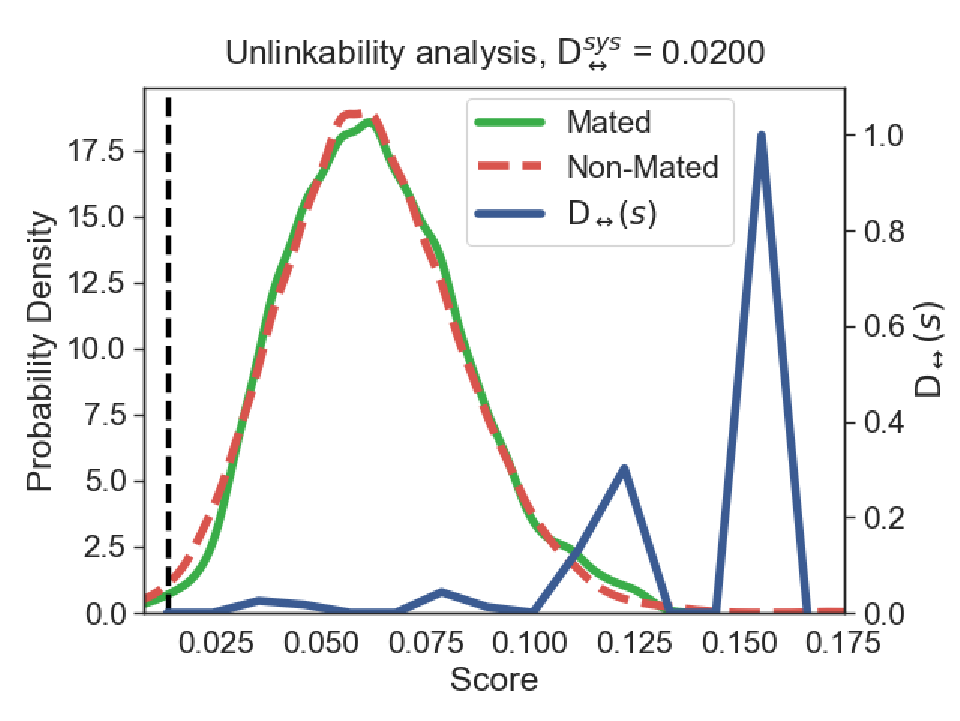}
  \caption{Unlinkability of BioDeephash on CASIA-IrisV4-Lamp}
\end{figure}
\subsection{Security Against Typical Attack Models}
In this section, we analyze the security of BioDeepHash in the context of nine specific attacks. The requirement, attack target, and attack feasibility of the nine kinds of attacks are shown in Table \ref{table:table1}. Each attack will be analyzed specifically next.
\begin{table*}[h!]
    \centering
    \caption{Security analysis of BioDeepHash against typical attacks}
    \renewcommand\arraystretch{1.5}
    \label{table:table1}
    \scalebox{1.1}{
    \begin{tabular}{c c c p{5.2cm}<{\centering}}
    \toprule
    \multirow{1}{*}{\textbf{Attack}} & \multirow{1}{*}{\textbf{Requirement}} & \multirow{1}{*}{\textbf{Attack target}} & \multirow{1}{*}{\textbf{Attack feasibility}}\\
    \midrule
    BF Attack & $\thicksim$ & Original biometric &Unrealistic complexity of BF attack\\
    \multirow{1}{*}{FA Attack} & \multirow{1}{*}{A large number of forged samples} & \multirow{1}{*}{Application system} & BioDeepHash has extremely low FAR\\
    ARM & BTP method don't satisfy Unlinkability & Original biometric & BioDeepHash satisfy Unlinkability\\  
    Statistical Attack & ECC used in BTP methods & Encryption key & ECC is not used in BioDeepHash\\
    Decodability Attack & ECC used in BTP methods & Protected template & ECC is not used in BioDeepHash\\
    Rainbow Table Attack & Suitable rainbow table & Encrypted hash value & Protected template is salted\\
    \multirow{2}{*}{Similarity Attack} & \multirow{2}{*}{BTP method satisfies distance-keeping property} & \multirow{2}{*}{Original biometric} & BioDeepHash don't satisfy ditance-keeping property\\  
    \multirow{2}{*}{MI Attack} & \multirow{2}{*}{target label or confidence vector} & \multirow{2}{*}{Original biometric} & It is impossible to enumerate the label or confidence vector corresponding to real users.\\
    \multirow{2}{*}{TI Attack} & \multirow{2}{*}{target label or confidence vector} & \multirow{2}{*}{Original biometric} & It is impossible to enumerate the label or confidence vector corresponding to real users.\\
    \bottomrule
    \end{tabular}
    }
\end{table*}
\subsubsection{Brute-Force Attack}\label{sec:bf}
The  BF attack \cite{ref42} refers to an attacker attempts to guess the biometric template by methodically enumerating all possible options. There are three potential points in the scheme that could suffer from BF attack:
\begin{enumerate}
    \item Training data \( x_u \)
    \item Output of the DFH model \( S_u \)
    \item Protected template stored in the database \( h(H_u) \)
\end{enumerate}
In BioDeepHash, we use $num_{G}$, the number of guesses an attacker needs to make to BF the target data, as a metric to measure the defensive effectiveness of our scheme against BF attack. The formula for calculating attack complexity is as follows:
\begin{equation}
    num_{G} = \prod_{i=1}^{dim} \prod_{j=1}^{size_{i}} R_{j}
\end{equation}
where $dim$ is the dimension of the target data, $size_{i}$ is the length of the $i$-th dimension of the target data, and $R_{j}$ is the range of the $j$-th pixel in $i$-th dimension of the target data. The training data is in the form of images. Assuming the image dimensions are \(a \times b\), the number of guesses towards the image $num_{G}$ is \(256^{a \times b}\). For the protected template processed by cryptographic hash function, the number of guesses towards the protected template $num_{G}$ is $2^{512}$. Obviously, both targets of attack mentioned above are unattainable. The best option for an attacker is to attempt enumeration of the output of the DFH model. In our scheme, the output of the DFH model is a binary string $S\in \{0,1\}^{L}$ of length \( L \). The number of guesses $num_{G}$ is \( 2^L \). Under the conditions described in section \ref{sec:exp:setup}, where the minimum secure length used in BioDeepHash is 96, an attack targeting the output of the DFH model is unfeasible.
\subsubsection{Decodability Attack}
The decodability attack \cite{ref26,ref27} is mainly used to distinguish whether two protected templates from different applications belong to the same user. This method targets the schemes that use ECC to determine whether the templates belong to the same user by decoding the dissimilar values of the two templates. In our framework, we do not use ECC but use deep hashing network to achieve error correction task, therefore, the attacker is unable to perform decodability attack on BioDeepHash.
\subsubsection{Statistical Attack}
The statistical attack \cite{ref51} is an attack scheme proposed by Rathgeb and Uhl against BTP methods using ECC. In this scheme the attacker obtains a large number of forged samples, decodes them, statistically obtains the frequency of occurrence of all the codewords, and selects the codeword with the largest frequency as the possible codeword. Statistical attack essentially exploits the situation where the probability of decoding a correct codeword is significantly higher than the average probability of decoding other codewords. Whereas, ECC is not used in BioDeepHash, the attacker is unable to perform statistical attack on BioDeepHash.
\subsubsection{Attack via Record Multiplicity (ARM)}
The ARM attack \cite{ref24} exploits multiple compromised protected templates to reconstruct original biometric data. In our scheme, the templates stored by the same user across different applications are differentiated by application-specific XOR strings and cryptographic hash functions. Since the XOR strings corresponding to the same user vary across different applications, the resulting protected templates after the XOR operation are quite different. After processing with the cryptographic hash function, protected templates undergo significant changes. When an attacker obtains compromised protected templates of the same user from multiple applications, it is completely impossible for the user to combine and utilize this information effectively. Furthermore, based on the unlinkability experiments shown in Fig. 4, BioDeepHash satisfies unlinkability. Consequently, attackers are unable to determine whether two protected templates belong to the same user. In summary, an ARM attack against BioDeepHash would degrade to a BF attack, which has already been proven infeasible.
\subsubsection{Similarity Attack}
The similarity attack \cite{ref14} fundamentally exploits the distance-keeping property of BTP methods. The distance-keeping property means that the distance between the protected template $g(x)$ and the guessing template $g(x^{*})$ remains constant relative to the distance between the original biometric data \( x \) and the guessing data \( x^{*} \). This property can be represented by the following formula:
\begin{equation}
    (1-\psi)d(x,x^{*})\le d(g(x),g(x^{*}))\le (1+\psi)d(x,x^{*})
\end{equation}
where $\psi$ denotes the error term, \(g(\cdot)\) is the BTP method. \(d(\cdot)\) is the distance function. Similarity attack solves the following optimization problem based on the aforementioned property:
\begin{equation}
    \operatorname*{argmin}_{x^{*}}d(g(x),g(x^{*})) 
\end{equation}
By solving this optimization problem, an approximation \( x' \) sufficiently close to the original biometric data \( x \) is obtained. In BioDeepHash, the final protected template is processed using cryptographic hashing. Due to the avalanche effect of cryptographic hashing, there is no distance-keeping properties between the stored protected template and the original biometric data. Therefore, the attacker is unable to perform similarity attack on BioDeepHash.
\subsubsection{False Acceptance (FA) Attack}
In the filed of BTP, the FA attack exploits the FAR of a BTP method to guess a fake sample that can be accepted by the system to pass verification \cite{ref43}. When the FAR of a BTP method is not 0, the attacker can expect to have a probability of passing the verification at each simulation attempt. In fact, FA attack has essentially relied on a similarity-based matching mechanism, when the similarity between the fake sample generated by the attacker and the real sample is greater than the matching threshold, it will pass the verification thus completing the FA attack. However, BioDeepHash has extremely low FAR values (e.g. 0\% for iris datasets, and $\leq$ 0.0002\% for facial datasets). We validate the ability of BioDeepHash to withstand FA attack through experiments. The specific experimental setup is to use samples from non-training users as fake samples to attack the BioDeepHash. In fact, this is similar to the open set experimental setup we used in section \ref{sec:exp:open}. Based on the results of the open-set test shown in Table \ref{table:table7}. We can conclude that BioDeepHash is effective in resisting FA attacks.

\subsubsection{Rainbow Table Attack}
The rainbow table attack \cite{ref52} is a type of attack against hash algorithms, where the attacker uses precomputed rainbow tables to crack hash values, thereby increasing attack efficiency. In BioDeepHash, we implement salting in the cancelable template generation module. This approach means that attackers would need to generate different rainbow tables for different users and applications, significantly increasing the complexity of the attack. As a result, our scheme is able to resist rainbow table attacks effectively.
\subsubsection{Model Inversion (MI) Attack}
The MI attack \cite{ref44} is a type of attack against NN, where the attack involves solving an optimization problem. In this attack, a randomly generated image is iteratively adjusted based on the output confidence of the NN until it resembles an image used in the training data. In this attack, the input is a target label or confidence vector, and the output is an image that corresponds to that label or confidence vector from the training data. In our model, the label can be considered as the $L$-dimensional binary code output by the DFH model. This means that the attacker must first determine a target label or confidence vector corresponding to a real user in order to retrieve the training image of that actual user through the attack. However, for an $L$-dimensional binary code, there are \(2^L\) possible combinations. Assuming there are \(N\) total users in an application, the probability that an attacker correctly identifies the real user's label is $\frac{N}{2^{L}}$. In BioDeepHash, the magnitude of \(N\) and \(2^L\) differs significantly, so $N$ can be ignored. We approximate that the probability of an attacker successfully guessing a user's label in each guess remains constant at $\frac{N}{2^{L}}$. Under such assumptions, the probability distribution for the attacker making \( k \) guesses can be modeled as a binomial distribution $B(k,\frac{N}{2^{L}})$. The formula for the probability of an attacker successfully guessing the enrolled user's label after \( k \) guesses is given by the following formula:
\begin{equation}
    P = 1-(\frac{2^{L}-N}{2^{L}} )^{k}
\end{equation}
For an attacker to achieve a feasible success probability, they must make a sufficiently large number of attack attempts. However, in practical scenarios, this is unachievable. In addition, Table \ref{table:table2} shows the number of enumerations \( k \) required for an attacker to achieve a success rate of 0.0001 under different datasets as well as different $L$ settings.
\begin{table}[h!]
  \centering
  \caption{$k$ under different dataset and $L$ settings}
  \label{table:table2}
  \scalebox{1.2}{
  \begin{tabular}{p{2.6cm}<{\centering} p{1cm}<{\centering} p{1cm}<{\centering} p{1cm}<{\centering} p{1cm}<{\centering}}
    \toprule
    \multicolumn{1}{c}{\multirow{2}{*}{\textbf{Dataset}}} & \multicolumn{3}{c}{\textit{\textbf{L}}} \\
    \cline{2-4}
    \specialrule{0em}{2pt}{2pt}
     & 96 & 120 & 144 \\
    \midrule
    Lamp($N$ = 804)           & 9.85e+21  & 1.65e+29 & 2.77e+36 \\
    Thousand($N$ = 2000)      & 3.96e+21  & 6.65e+28 & 1.12e+36 \\
    Yale face($N$ = 38)      & 2.09e+23  & 3.50e+30 & 5.87e+37 \\
    YouTube face($N$ = 1595)   & 4.97e+21  & 8.33e+28 & 1.40e+36 \\
    \bottomrule
  \end{tabular}
  }
\end{table}
\subsubsection{Template Inversion Attack}\label{sec:TIA}
In TI attack, the attacker gains access to protected templates stored in the database and attempts to invert these templates to reconstruct the biometric data. According to the discussion in Section \ref{sec:irre}, to perform TI attack against BioDeepHash, the first issue to consider is how to obtain the model output corresponding to the training data. A simple strategy is to enumerate the model outputs, however this strategy has been discussed as infeasible in Section \ref{sec:irre}. In addition to the enumeration strategy, there exists a guessing strategy \cite{ref61}, which obtains the model output corresponding to some of the training data by using a large number of images of non-registered users as inputs to the model, and then exploits the model's false acceptance case (when the model recognizes an illegal image as a legitimate user, the attacker can use the illegal image to obtain the output corresponding to the registered user in the model). In the open set experiments, we can see that the FAR of BioDeepHash for illegal users is 0, which means that it is infeasible to obtain the model output corresponding to the training data by guessing. In summary the TI attack is also not feasible for BioDeepHash.

%% file: sec/5_exp.tex
\section{Experiment}
In this section, we introduce the datasets used and the specific settings of our experiments. Subsequently, we present the performance results of BioDeepHash. Finally, we compare BioDeepHash with existing BTP methods.
\subsection{Experimental Setup}\label{sec:exp:setup}
\subsubsection{Dataset}
To validate the universality of our framework, we conducted experiments on two commonly used biometric data types: facial data and iris data. For facial data, we used the YouTube faces dataset and the Extended Yale B dataset, and for iris data, we used the CASIA-Iris-Lamp dataset and the CASIA-Iris-Thousand dataset. Next, we provide a brief introduction to each of these four datasets.
\begin{itemize}
    \item \textbf{YouTube faces dataset (YouTube face)}: The YouTube faces dataset \cite{ref45} is a facial video dataset that contains 3453 videos from 1595 users. We selected 40 images per user for training and 5 images for testing with reference to \cite{ref57}.
    \item \textbf{Extended Yale B dataset (Yale face)}: The Yale face dataset \cite{ref46} consists of 2432 images with different lighting variations from 38 users. Following the settings described in \cite{ref33}, we used the cropped version of the dataset, selecting 10 images per user for training and the remaining images for testing. To mitigate the impact of lighting variations, we applied lighting normalization to all images using the method described in \cite{ref47}.
    \item \textbf{CASIA-Iris-Lamp dataset (Lamp)}: The Lamp dataset contains images from 411 users, totally 16212 images. During the image collection process, a lamp was turned on and off, creating different lighting conditions. These conditions can cause elastic deformations in iris textures, leading to significant intra-class variations and making testing on this dataset more challenging. For this dataset, we treat the left and right irises of the same user as separate users with reference to \cite{ref22}. We select 4 images per user for testing, with the remaining images for training.
    \item \textbf{CASIA-Iris-Thousand dataset (Thousand)}: The Thousand dataset contains 20000 iris images from 1000 users. In this dataset, many users are wearing glasses, which causes significant intra-class variability. This dataset is the first publicly available iris dataset with a thousand individuals, offering a rich source of data. For this dataset, we treat the left and right irises of the same user as separate users with reference to \cite{ref22}. We select 2 images per user for testing, with the remaining images for training.
\end{itemize}

\subsubsection{Network Configuration and Evaluation Metrics}
As introduced in Section \ref{sec:method:dfh}, our DFH model primarily consists of feature extraction layers and hash layer. The specific architecture of the network is shown in Fig. 3 and is composed of 8 convolutional layers and 2 fully connected layers. Each convolutional layer is followed by a $2\times 2$ pooling layer. Batch normalization and ReLU activation functions are applied after all convolutional layers and the first fully connected layer. For training the DFH model, we use an Adam optimizer with a learning rate of 0.005. The batch size for all datasets is set to 512, and the training is conducted over 350 epochs. For the Thousand dataset, the parameter settings are fixed at $\sigma$ = 8, $\eta$ = 0.02. For the other three datasets, the parameter settings are fixed at $\sigma$ = 16, $\eta$ = 0.02. Additionally, for comparison with \cite{ref33}, the Yale face dataset uses the second matching strategy, while the rest of the datasets uses the first matching strategy.

In our experiments, we evaluated BioDeepHash under five different lengths $L$ of DFH model output, using FAR and GAR as evaluation metrics. FAR denotes the proportion of illegal attempts that were incorrectly accepted, and GAR denotes the proportion of legal attempts that were correctly accepted.

\begin{figure*}
\centering
	\begin{minipage}{0.49\linewidth}
		\centering
		\includegraphics[width=0.9\linewidth]{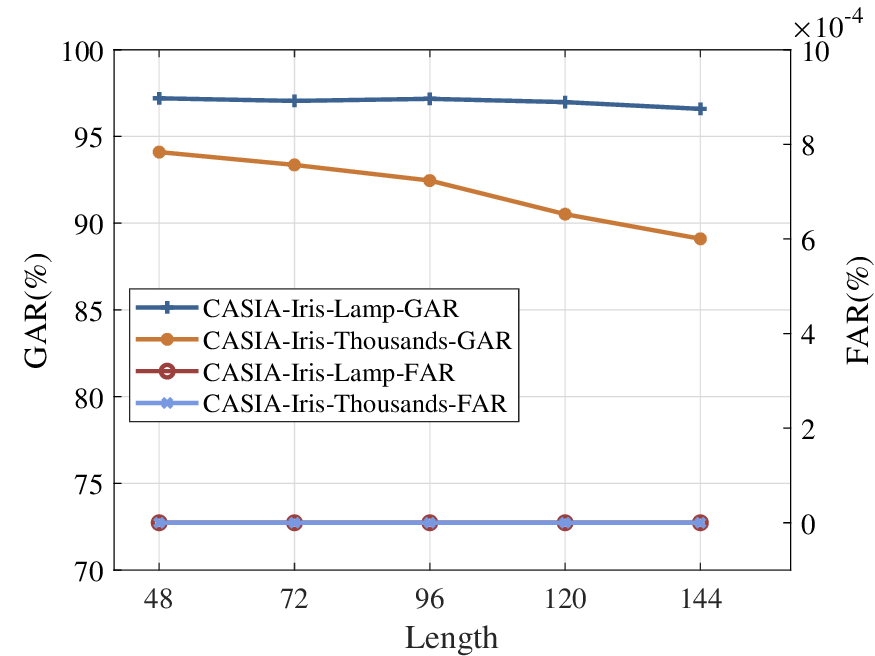}
        \captionsetup{justification=raggedright}
		\caption{GAR and FAR ($\%$) for different $L$ on Lamp and Thousand}
        \label{fig:fig5}
	\end{minipage}
	\begin{minipage}{0.49\linewidth}
		\centering
		\includegraphics[width=0.9\linewidth]{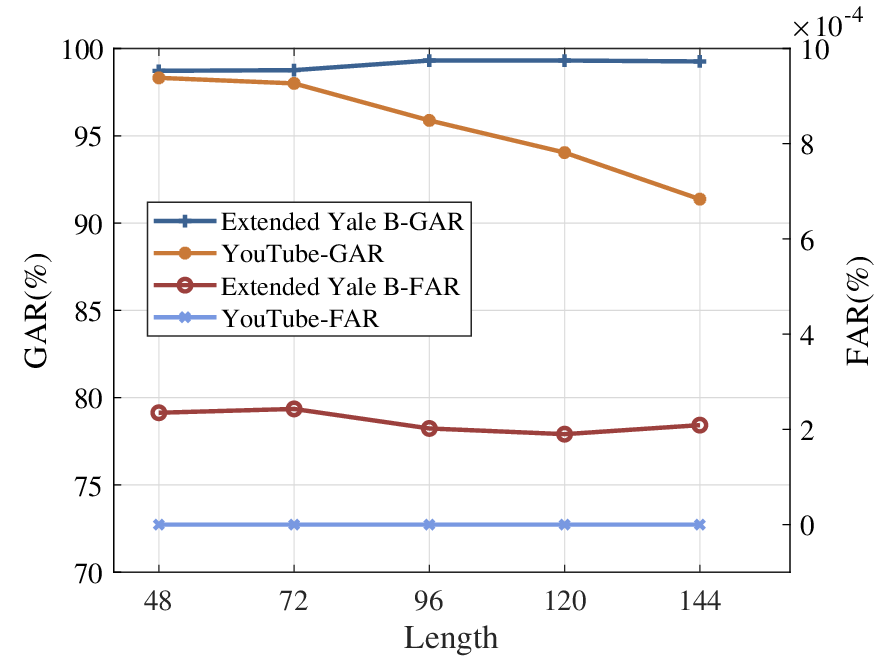}
        \captionsetup{justification=raggedright}
		\caption{GAR and FAR ($\%$) for different $L$ on Yale face and YouTube face}
        \label{fig:fig6}
	\end{minipage}
\end{figure*}

\subsection{Performance Experiments}
\subsubsection{Performance Evaluation on Different Datasets}\label{sec:exp:perform}
As shown in Fig. \ref{fig:fig5} and Fig. \ref{fig:fig6}, BioDeepHash achieves high performance as well as extremely low FAR on four datasets. The high performance of BioDeepHash comes from the powerful processing of images by deep hashing models. Many existing deep learning-based BTP schemes similarly exhibit high performance, but BioDeepHash has another advantage: extremely low FAR. Based on our analysis in irreversibility as well as MI attack, FA attack, and TI attack, we can find that many attack methods rely on false acceptance of BTP methods to steal private information. This suggests that extremely low FAR is important to ensure data privacy. We attribute the very low FAR to the use of the DFH model and the precision matching strategy. DFH model is trained on a targeted sample of registered users, allowing the model to more accurately identify ``non-participating users'' (i.e., illegitimate users), thus guaranteeing a low FAR. Similarly, the strong image processing capability of the deep learning model allows the model to distinguish the categories between training samples, reducing misclassification (misclassification refers to the conversion of images of different users into the same code), which in turn reduces the FAR. In addition, the precise matching strategy allows our scheme to reject users with even a one-place difference, which further reduces the FAR.

In addition, Fig. \ref{fig:fig5} and Fig. \ref{fig:fig6} show that the GAR of our scheme decreases as the output length of the DFH model increases on the Lamp, Thousand, and Youtube face datasets. This pattern is also obeyed in the Yale face dataset where $L$ is greater than 96. However, the length plays a crucial role in security. In this section of experiments, we choose five lengths in order to investigate the effect of $L$ on the performance. In fact, $L$ can only guarantee the security when it is larger than 96, so in the subsequent experiments, we only choose 96, 120, and 144. 

\begin{table*}[h!]
    \centering
    \caption{Comparison of BioDeepHash with state-of-the-art methods}
    \label{table:table5}
    \scalebox{1.3}{
    \begin{tabular}{c c c c c c c}
    \toprule
    \multirow{1}{*}{\textbf{Data type}} & \multirow{1}{*}{\textbf{Method}} & \multirow{1}{*}{\textbf{Dataset}} & \multirow{1}{*}{\textbf{GAR@FAR} ($\%$)} & \multirow{1}{*}{\textbf{Without ECC}} & \multirow{1}{*}{\textbf{Revocability}} & \multirow{1}{*}{\textbf{Cryptographic hash}}\\
    \midrule
    \multirow{8}{*}{Iris} & \multirow{2}{*}{Zhao et al.\cite{ref53}} & Lamp & 97.01@0.01 & \multirow{2}{*}{$\checkmark$} & \multirow{2}{*}{$\checkmark$} & \multirow{2}{*}{$\times$} \\
    &  & Thousand & 88.95@0.01 & \\
    \cline{3-4}
    & \multirow{2}{*}{Lee et al.\cite{ref15}} & Lamp & $\thicksim$ & \multirow{2}{*}{$\checkmark$} & \multirow{2}{*}{$\checkmark$} & \multirow{2}{*}{$\times$} \\
    &  & Thousand & 84.68@0.01 \\
    \cline{3-4}
    & \multirow{2}{*}{Zhao et al.\cite{ref7}} & Lamp & 79.07@0.01 & \multirow{2}{*}{$\checkmark$} & \multirow{2}{*}{$\checkmark$} & \multirow{2}{*}{$\times$} \\
    &  & Thousand & $\thicksim$ \\
    \cline{3-4}
    & \multirow{2}{*}{Othman et al.\cite{ref54}} & Lamp & 79.95@0.01 & \multirow{2}{*}{$\checkmark$} & \multirow{2}{*}{$\times$} & \multirow{2}{*}{$\times$} \\
    &  & Thousand & 78.50@0.01 \\
    \cline{3-4}
    & \multirow{2}{*}{BioDeepHash} & Lamp & $\textbf{97.17}$@$\textbf{0.00}$ & \multirow{2}{*}{$\checkmark$} & \multirow{2}{*}{$\checkmark$} & \multirow{2}{*}{$\checkmark$} \\
    &  & Thousand & $\textbf{92.46}$@$\textbf{0.00}$ \\
    \midrule
    \multirow{10}{*}{Face} & \multirow{2}{*}{Pandey et al.\cite{ref32}} & YouTube face & $\thicksim$ & \multirow{2}{*}{$\checkmark$} & \multirow{2}{*}{$\times$} & \multirow{2}{*}{$\checkmark$} \\
    &  & Yale face & 96.74@0.00 \\
    \cline{3-4}
    & \multirow{2}{*}{Chen et al.\cite{ref33}} & YouTube face & $\thicksim$ & \multirow{2}{*}{$\times$} & \multirow{2}{*}{$\times$} & \multirow{2}{*}{$\checkmark$} \\
    &  & Yale face & 99.16@1.00 \\
    \cline{3-4}
    & \multirow{2}{*}{Jang et al.\cite{ref57}} & YouTube face & $\textbf{97.04}$@$\textbf{0.10}$ & \multirow{2}{*}{$\checkmark$} & \multirow{2}{*}{$\checkmark$} & \multirow{2}{*}{$\times$} \\
    &  & Yale face & $\thicksim$ \\
    \cline{3-4}
    & \multirow{2}{*}{Pinto et al.\cite{ref58}} & YouTube face & 84.96@0.10 & \multirow{2}{*}{$\checkmark$} & \multirow{2}{*}{$\checkmark$} & \multirow{2}{*}{$\times$} \\
    &  & Yale face & $\thicksim$ \\
    \cline{3-4}
    & \multirow{2}{*}{BioDeepHash} & YouTube face & $\textbf{95.88}$@$\textbf{2.02e-4}$ & \multirow{2}{*}{$\checkmark$} & \multirow{2}{*}{$\checkmark$} & \multirow{2}{*}{$\checkmark$} \\
    &  & Yale face & $\textbf{99.31}$@$\textbf{0.00}$ \\
    \bottomrule
    \end{tabular}
    }
\end{table*}

\begin{table}[!ht]
    \centering
    \caption{GAR@FAR ($\%$) for different $\sigma^{2}$ and $\eta$ on Lamp}
    \label{table:table4}
    \scalebox{1.15}{
    \begin{tabular}{c c c c c c}
        \toprule
        
        \multirow{2}{*}{$ \sigma^{2} $} & \multirow{2}{*}{{\textit{\textbf{L}}}} & \multicolumn{4}{c}{$\mathbf{ \eta} $}   
        \\  \cline{3-6}
        ~ & ~ & 0.0025 & 0.005 & 0.01 & 0.02  \\  \cline{1-6}
        \specialrule{0em}{2pt}{2pt}
        \multirow{3}{*}{2} & 96bit & 7.42@0 & 20.02@0 & 44.67@0 & 62.96@0  \\ 
         & 120bit & 2.39@0 & 12.70@0 & 43.65@0 & 62.66@0  \\ 
         & 144bit & 2.28@0 & 12.13@0 & 42.05@0 & 62.53@0  \\  
        \midrule
        \multirow{3}{*}{4} & 96bit & 58.73@0 & 79.58@0 & 85.11@0 & 89.26@0  \\ 
        ~ & 120bit & 42.30@0 & 74.31@0 & 77.53@0 & 83.10@0  \\ 
        ~ & 144bit & 32.53@0 & 56.32@0 & 71.36@0 & 82.06@0  \\ 
        \midrule
        \multirow{3}{*}{8} & 96bit & 73.16@0 & 89.46@0 & 94.05@0 & 94.72@0  \\ 
        ~ & 120bit & 73.60@0 & 86.95@0 & 92.45@0 & 93.32@0  \\ 
        ~ & 144bit & 70.26@0 & 83.51@0 & 91.75@0 & 93.29@0  \\ 
        \midrule
        \multirow{3}{*}{16} & 96bit & 77.47@0 & 93.41@0 & 96.64@0 & 97.17@0  \\ 
        ~ & 120bit & 77.87@0 & 92.16@0 & 96.44@0 & 96.98@0  \\ 
        ~ & 144bit & 79.24@0 & 93.08@0 & 96.14@0 & 96.59@0  \\
        \bottomrule
    \end{tabular}
    }
\end{table}

\subsubsection{Parameter Variation Experiments}
To investigate the impact of parameters on BioDeepHash, we conducted parameter variation experiments. The variable parameters in our framework mainly include \( L \), \( \eta \) and \( \sigma^{2} \). The parameter variation experiments were conducted on the Lamp dataset. The \( \sigma^{2} \) was set within $\{2,4,8,16\}$, \( \eta \) was set within $\{0.0025,0.005,0.01,0.02\}$, \( L \) was set within $\{96,120,144\}$. The results are displayed in Table \ref{table:table4}. As shown, the GAR gradually increases as \( \sigma^{2} \) increases under the same \( L \), \( \eta \) settings. According to conclusions in \cite{ref38}, a larger \( \sigma \) reduces intra-class variations, enabling the model to better recognize samples within the same class. Additionally, when \( \sigma^{2} \) exceeds 32, GAR is close to 0, so we did not show it in Table \ref{table:table4}. We think this is due to \( \sigma^{2} \) , which is used as a global scaling factor to control inter-class disparity, being too large. This causes class gaps to become too small, preventing the model from learning the differences within classes in the feature space. We can see that as \( \eta \) increases under the same \( L \), \( \sigma^{2} \) settings, the GAR gradually increases. This occurs because a decrease in \( \eta \) reduces the proportion of quantization loss, resulting in hash codes that derive less information from the labels. Consequently, images of the same user struggle to generate a stable code. In the ablation study in Section \ref{sec:exp:adla}, we can more clearly see the importance of quantization loss in enabling images from the same user to generate perfectly a stable code.

\subsection{Comparison with State-of-the-Art Methods}
We compare BioDeepHash with existing approaches using facial and iris data. Performance is evaluated through GAR@FAR metrics. The security of these methods is assessed based on the use of ECC, satisfaction of revocability, and employment of cryptographic hashing. The results in Table \ref{table:table5}, demonstrate that BioDeepHash outperforms other methods on the Thousand and Lamp datasets for iris data, achieving the best results. For facial data, BioDeepHash is tested on the Yale and YouTube face datasets. It outperforms other methods on the Yale dataset, and although the GAR is slightly lower than \cite{ref57} on the Youtube dataset, our method achieves a lower FAR compared to the scheme in \cite{ref57}. From a security perspective, BioDeepHash does not use ECC, ensures revocability and employs cryptographic hashing to enhance security. None of the other methods simultaneously meet all these three criteria.

\subsection{Ablation Experiments}\label{sec:exp:adla}
In DFH model we utilized, multiple loss functions are combined to optimize performance. To investigate the impact of these loss functions on DFH model's effectiveness, we conduct ablation experiments. These experiments help identify the contribution of each loss component to the overall performance of the DFH model. According to Formula 11, the loss functions we employed can be divided into: $L1$, $L2$, and the quantization loss $L3$. We conducted three sets of experiments on the Lamp dataset to explore their impact: $L1$ indicates that we only use the loss from Formula 7. $L1+L3$ indicates that we combine the loss from Formula 7 with the quantization loss. $L1+L2+L3$ indicates that we apply all loss functions.

\begin{table}[htbp]
    \centering
    \caption{GAR@FAR ($\%$) results of ablation experiments}
    \label{table:table6}
    \scalebox{1.2}{
    \begin{tabular}{c c c c c c}
    \toprule
    \multicolumn{3}{c}{\textbf{Loss function}} & \multicolumn{3}{c}{\textit{\textbf{L}}} \\ \cline{4-6}
    \specialrule{0em}{2pt}{2pt}
    $L1$ & $L2$ & $L3$ & 96 & 120 & 144 \\
    \midrule
    $\checkmark$ & &           & 0.03@0  & 0.01@0 & 0.00@0 \\
    $\checkmark$ & & $\checkmark$       & 96.42@0  & 96.36@0 & 95.73@0 \\
    $\checkmark$ & $\checkmark$ & $\checkmark$ & 97.17@0  & 96.98@0 & 96.59@0 \\
    \bottomrule
    \end{tabular}
    }
\end{table}

The results are shown in Table \ref{table:table6}. It can be observed that the GAR values gradually increase as more loss functions are combined. When only the $L1$ loss is used, the GAR is close to 0 because our model can only map images of the same user to similar codes, not to a stable code. When both $L1$ and $L3$ are used, there is a significant improvement in model performance. This suggests that quantization loss $L3$ plays a very important role in getting stable code. The best performance is achieved when $L1$, $L2$, and $L3$ are used together, demonstrating the effectiveness of the combined loss functions on enhancing the model's ability to accurately map images from same user to a stable code.

\subsection{Open Set Experiment}\label{sec:exp:open}
In order to verify the rejection ability of BioDeepHash for unregistered users, we set up two types of experiments to simulate user authentication in the open set condition: Lamp(Thousand) and Thousand(Lamp). The setup of Lamp(Thousand) is: 200 classes are randomly selected in Lamp dataset, 4 images are randomly selected each class for testing, the rest of the images are used for training, and then 200 classes are randomly selected from the remaining classes, where 4 images are selected for each class as unknown samples for testing. Thousand(Lamp) is setup as: 400 classes are randomly selected in Thousand dataset, for each class randomly select 2 images for testing, the rest of the images are used for training, and then randomly select 400 classes from the remaining classes, where for each class select 2 images as unknown samples for testing. We using GAR and FAR as evaluation metrics. The results of the open-set test are shown in Table \ref{table:table7}. The FAR is 0 for both settings. This demonstrates that BioDeepHash is able to correctly reject unknown users under the open-set setting. It demonstrates that BioDeepHash is effective in preventing users from stealing the model output of training data through a large number of unknown samples when resisting TI attacks. It is worth noting that the GAR obtained here is slightly different from that in the performance experiment, which is due to the fact that some of the classes were randomly selected in the open-set experiment while all the classes were used in the performance experiment.

\begin{table}[h!]
  \centering
  \caption{GAR@FAR ($\%$) results of open set experiments}
  \label{table:table7}
  \scalebox{1.2}{
  \begin{tabular}{c c c c c}
    \toprule
    \multicolumn{1}{c}{\multirow{2}{*}{\textbf{Open set condition}}} & \multicolumn{3}{c}{\textit{\textbf{L}}} \\
    \cline{2-4}
    \specialrule{0em}{2pt}{2pt}
     & 96 & 120 & 144 \\
    \midrule
    Lamp(Thousand)      & 96.63@0  & 96.50@0 & 95.12@0 \\
    Thousand(Lamp)      & 93.55@0  & 93.16@0 & 93.30@0 \\
    \bottomrule
  \end{tabular}
  }
\end{table}

%% file: sec/6_conclusion.tex
\section{Conclusion}
In this paper, we propose a BTP framework called BioDeepHash that combines deep hashing and cryptographic hashing. In BioDeepHash, different images from the same user are mapped to a stable code using a deep hashing model. Then we use a cancelable template generation module to convert the application-specific XOR string and the stable codes into cancelable templates. Finally the cancelable templates are processed using cryptographic hashing to get the final protected templates. Experiments on facial and iris datasets show that BioDeepHash can achieve promising results on Thousand dataset, Lamp dataset, YouTube dataset and Yale facial dataset. The FAR of BioDeepHash in the open-set environment is extremely low. In addition, experimental and theoretical analyses show that BioDeepHash satisfies irreversibility, unlinkability, and revocability.

In the future work, we will consider optimizing our deep hashing model to further improve the recognition performance. In addition, we also consider using multiple deep hashing models and stitching the outputs of multiple deeping hash models to increase the length of binary codes and enhance the security.